\documentclass[aps,pra,showpacs,twoside,twocolumn,10pt]{revtex4-2}
\usepackage[colorlinks=true, citecolor=blue, urlcolor=blue, linkcolor = blue ]{hyperref}
\usepackage{epsfig,newlfont,amssymb,amsfonts,amsmath,bm,subfigure,palatino,mathtools,amsthm,braket,times,soul,enumitem,color}
\usepackage[normalem]{ulem}
\newcommand{\stkout}[1]{\ifmmode\text{\sout{\ensuremath{#1}}}\else\sout{#1}\fi}
\usepackage[T1]{fontenc}
\usepackage[normalem]{ulem}
\usepackage{graphicx}
\usepackage{placeins}
\usepackage{hyperref}
\usepackage{multirow}
\usepackage{xcolor}
\usepackage{hhline,booktabs}

\begin{document}

\preprint{}

\title{Enhanced sensing of a weak Stark field under the influence of Aubry-Andr{\'e}-Harper criticality}

\author{Ayan Sahoo}
\affiliation{Harish-Chandra Research Institute, A CI of Homi Bhabha National Institute,
Chhatnag Road, Jhunsi, Allahabad 211 019, India}

\author{Debraj Rakshit}
\affiliation{Harish-Chandra Research Institute, A CI of Homi Bhabha National Institute,
Chhatnag Road, Jhunsi, Allahabad 211 019, India}


\begin{abstract}
The localization transition can be exploited as a resource for achieving quantum-enhanced sensitivity in parameter estimation. We demonstrate that by employing different classes of localization inducing potentials, one can significantly enhance the precision of parameter estimation. Specifically, we focus on the precision measurement of the Stark strength parameter encoded in the low- and high-energy eigenstates of a one-dimensional fermionic lattice under the influence of Aubry-Andr{\'e}-Harper localization-delocalization transition. For the ground state, we consider the single-particle system, in addition to the system at half filling. Our work reveals that Quantum Fisher Information (QFI) offers superior scaling with respect to the system size compared to the pure stark case, leading to a better parameter estimation. However, experimental measurement of the QFI based on fidelity in a multibody system is a significant challenge. To address this, we suggest experimentally relevant operators that can be utilized to achieve precision surpassing the Heisenberg Limit (HL) or can even saturate the QFI scaling. These operators, relevant for practical experimental setups, provide a feasible pathway to harness the advantages offered by the localization-delocalization transition by exploiting two distinct localizing potentials for quantum-enhanced parameter estimation. In addition, to investigate the robustness of the proposed quantum sensors, we add noise in the form of thermal fluctuations, and demonstrate that the Fisher information of the thermal states can saturate to HL. In principle, this work demonstrates how introducing additional quantum-criticality-inducing control parameters could be utilized to enhance quantum sensitivity in the estimation of an unknown target parameter, and hence promises applications in wider contexts, e.g., in hybrid systems involving other classes of localization inducing potential, such as Anderson localization, and even in systems involving other types of quantum criticalities, such as second-order and topological phase transitions. 

\end{abstract}

\maketitle
\section{Introduction}  The ultimate limit of precision in parameter estimation is given by the Cram\'{e}r-Rao bound \cite{PhysRevLett.72.3439,RevModPhys.89.035002,Cramer1946,1969JSP.....1..231H}. For an unknown parameter, $h$, the standard deviation of parameter $\delta h$ is bounded by the relation,  $\delta h \geq 1/\sqrt{MF_Q}$, where $M$ is the number of repetitions of the sensing protocol and $F_Q$ is the quantum Fisher information (QFI). Classically, the QFI scales with $L$ independent qubit as $F_Q \sim L$, known as a standard quantum limit (SQL), whereas it can reach the Heisenberg limit (HL), i.e., $F_Q \sim L^2$, for certain quantum mechanical systems \cite{giovannetti2004quantum,PhysRevLett.96.010401,PhysRevX.8.021022,PhysRevA.78.042105,PhysRevLett.99.095701,PhysRevA.88.021801,PhysRevLett.124.120504,PhysRevLett.127.080504,Gietka2021adiabaticcritical,PhysRevLett.126.010502,SaladoMeja2021,PhysRevLett.126.200501,PhysRevA.101.043609,PhysRevLett.121.020402,PhysRevE.74.031123,PhysRevE.76.022101,PhysRevLett.99.100603,RevModPhys.90.035006,PRXQuantum.3.010354,PhysRevX.13.031012,agarwal2025quantum,Montenegro2025}. There are some special kind of entangled states, such as the Greenberger-Horne-Zeilinger (GHZ) state that can be exploited for achieving the HL. However, GHZ-based quantum sensors are susceptible to decoherence. Recently, various kinds of quantum many-body probes have been proposed for achieving quantum-enhanced sensitivity by harnessing the different types of quantum phenomena. Quantum many-Body (QMB) systems, such as second-order, localization-delocalization, and topological phase transitions have been identified as a quantum resource for parameter sensing \cite{PhysRevLett.110.146404,PhysRevLett.110.176403,PhysRevResearch.3.013148,PhysRevX.8.021022,PhysRevA.78.042105,PhysRevA.93.022103,Garbe_2022}. Other types of many-body probes that have been reported include quantum scars \cite{PhysRevB.107.035123,yoshinaga2022quantum,PhysRevLett.126.210601,PRXQuantum.2.020330} and Floquet driving \cite{PhysRevLett.127.080504}. Generally, the precision of these probes can reach the HL. Moreover, sensing protocols with super-Heisenberg scaling have also been proposed recently on the QMB platform \cite{PhysRevB.77.245109,PhysRevX.8.021022,PhysRevA.99.042117,PhysRevLett.98.090401,PhysRevLett.100.220501}. {\color{black}{There is some nuance in the behaviors that constitute a resource in the critical parameter-estimation sensing: it is the confluence of diverging susceptibility, enhanced responses to small perturbations, and the generation of long-range correlations, symmetry breaking, and, in some circumstances, gap closing, which can lead to an enhancement \cite{Montenegro2025,mihailescu2025critical,mondal2024multicritical}.}}

QMB systems hosting distinct quantum phases can be used as a sensing probe. {\color{black}{One may consider estimating a parameter that is encoded in the ground state or in an excited state of a system. These types of sensors may achieve the HL at the criticality \cite{Gu_2014,PhysRevLett.100.100501,PhysRevA.79.012305}. In order to estimate an unknown parameter governing the quantum phase transition in these types of sensors, we need to prepare a state and then adiabatically tune the parameter to be estimated near criticality. This approach is referred as adiabatic sensing. In the critical region, the system becomes sensitive to small change to the parameter, resulting in a divergent quantum Fisher information.}}  As an example, due to a self-dual symmetry, the Aubry-Andr{\'e}-Harper (AAH) model shows localization-delocalization transition at a finite strength of the potential amplitude. This transition characterized by a quasiperiodic potential can be used as a resource for achieving quantum-enhanced parameter sensing that attains the HL \cite{PhysRevA.109.L030601}. Recently, there have been proposals for quantum sensors using Stark localization. Interestingly, one can achieve the super Heisenberg limit via Stark localization transition \cite{PhysRevLett.131.010801}. QMB sensors have various applications in quantum technology \cite{appel,Louchet2010, RevModPhys.90.035005,PhysRevLett.79.3865, PhysRevX.6.041044,PhysRevX.10.031003,PhysRevLett.131.220801,10.21468/SciPostPhys.13.4.077,Mihailescu_2024,Yousefjani2024,mourik2012signatures,Deng2012,Das2012,Fulga_2013,Nadj-Perge2014,Xu2016,Xing2018,Dvir2023,manshouri2024quantum}. At the critical point or transition point the fidelity susceptibility scales with system size $L$ as $L^{2/d\nu}$, where $d$ represents the spatial dimensionality of the system, and $\nu$ is the critical exponent associated with the localization length, $\zeta$, near criticality ($h_c$), such that $\zeta \sim |h - h_c|^{-\nu}$. At the transition point, the precision of the unknown parameter $h$ scales with system size as $L^{2/\nu}$ \cite{PhysRevB.86.245424, PhysRevB.101.174203}. {\color{black}{Implementing the optimal measurement that saturates the QFI is often challenging in practice. Therefore, it is necessary to identify an operator that is both theoretically meaningful and experimentally accessible.}} For operators, one needs to compute operator-based Fisher information (OFI), which is basically defined via signal-to-noise ratio. It can be shown that OFI is bounded by QFI \cite{pezze2019adiabatic}. In principle, the OFI saturates QFI when optimized over all possible operators.

The AAH model, characterized as a tight-binding model with nearest-neighbor hopping and a quasiperiodic onsite potential, undergoes a localization-delocalization phase transition when the onsite potential's strength reaches twice the value of the hopping term \cite{aah_model, Hetenyi2024, Hetenyi2025}. The experimental evidence of this transition for the AAH model has been reported in ultracold atom setup \cite{doi:10.1126/science.aaa7432,PhysRevLett.90.055501,Modugno_2010} and in photonic crystals \cite{PhysRevLett.103.013901,PhysRevLett.109.106402,PhysRevLett.110.076403,PhysRevB.91.064201}. On the other hand, in the tight-binding model, if a linear gradient field is induced across the lattice instead of quasiperiodic potential, the system also shows a localization-delocalization transition in the limit of zero field strength. This is the case of well-known Stark localization \cite{PhysRev.117.432, PhysRevB.8.5579, jiang2023stark, PhysRevB.106.134207}. In the context of the AAH model and the Stark model, the QFI exhibits distinct scaling behaviors at their respective criticalities. Specifically, at the criticality of the AAH model, the QFI scales with system size as $L^2$ for single particle as well as half-filled probe, which reaches the HL. On the other hand, at the criticality of the Stark model, the QFI scales as $L^{5.9}$ for single particle probe and $L^{4.1}$ for a half-filled probe, surpassing the Heisenberg limit \cite{PhysRevLett.131.010801}. Numerous studies have explored these critical phenomena individually in the presence of weak interactions \cite{PhysRevLett.113.045304,PhysRevB.87.134202}. Recent studies have investigated the characteristic properties of fundamental physical quantities, such as localization length, inverse participation ratio, and energy gap, by examining systems in the presence of these two different types of potentials. These studies reveal that the scaling exponents associated with the localization length, energy gap, and other related quantities are markedly different from the pure cases  \cite{sahoo2024stark}. In this work, we investigate the advantage of interplay between these two potentials in the quantum sensing of a weak Stark field while initiating the system at the AAH criticality. Our results demonstrate that the precision of the parameter encoded in the Stark field potential can be significantly enhanced when measuring the parameter $h$ in the presence of the AAH criticality. This suggests a novel approach to improve parameter estimation in quantum systems by leveraging the critical properties of different classes of quantum phase transitions, in general. 
 
The introduction is presented in Sec.~I. Section II presents the fidelity-based definitions of QFI and OFI. The system is introduced in Sec.~III. Sections IV and V examine the behavior of QFI and OFI, as well as their scaling with system size, for a single-particle probe in the ground and mid-spectrum states, respectively. Section VI explores the behavior of the QFI and OFI for the thermal states. Section VII investigates QFI, OFIs, and their scaling for a many-body half-filled probe. Finally, Sec.~VIII provides a Conclusion.

\section{parameter estimation} If an unknown parameter $h$ is encoded in a quantum state $|\psi(h)\rangle$, the fluctuation of the parameter near the $h$ is captured by the fidelity susceptibility $\chi_Q$ which is defined by
\begin{equation}
 \chi_Q = -\lim_{\delta h \to 0}\frac{\partial^2 {\cal F}_Q}{\partial (\delta h)^2},
 \label{eq:fidelitysucc}
\end{equation}
where ${\cal F}_Q =  |\langle \psi(h)|\psi(h+\delta h)\rangle|$ is the distance between two nearby states  $|\psi(h)\rangle$ and $|\psi(h+\delta h)\rangle$. The QFI is related to fidelity susceptibility as ${\cal F} = 4\chi_Q$ \cite{PhysRevA.78.042105}. {\color{black}{Estimating an unknown parameter requires measuring an observable whose expectation value depends on that parameter. For an operator $\hat O$, $\langle \hat O \rangle = \text{Tr}[\hat O \rho(h)]$ is the function of $h$.}} where $\rho(h)$ is the quantum state of the sensor. The sensitivity of $h$ is quantified by the error propagation formula, which is essentially the signal-to-noise ratio, defined as
\begin{equation}
 F_O = \lim_{\delta h \to 0}  \frac{\Big(\frac{d\langle {\hat{O}}\rangle}{d(\delta h)}\Big)^2}{\text{Var}(\hat{O})},
 \label{eq:OFI}
\end{equation}
where $\text{Var}(\hat{O})$ is defined as, $\text{Var}(\hat{O}) = \langle {\hat{O}}^2 \rangle -\langle {\hat{O}} \rangle^2$. 
The quantum Cram{\'e}r-Rao bound~\cite{Cramer1946,1969JSP.....1..231H,pezze2019adiabatic} provides the bound on the uncertainty for any observable estimation: $F_O(h, \hat{O}) \le F_Q(h)$.

\section{System}  We consider a one-dimensional lattice with nearest neighbor hopping in the presence of a quasiperiodically modulated potential (Aubry–Andr{\'e}) and a linear gradient (Stark) field, referred to as the Aubry–Andr{\'e}–Stark (AAS) model. {\color{black}The generic form of Hamiltonian is $\hat{H} = \hat{H_1}  + h \hat{H_2}$. The explicit form of the Hamiltonian is given by,
\begin{align}
\label{eq:Ham}
 \hat{H_1} &= - \sum_{i}^{L-1}(\hat{c}^{\dagger}_{i}\hat{c}_{i+1}+\mathrm{H.c.}) 
 \nonumber + V \sum_{i}^{L-1}  \cos[2\pi (i \omega + \phi)] \hat{c}^{\dagger}_{i}\hat{c}_{i},  \\
\hat{H_2} &= h \sum_{i}^{L-1} i \hat{c}^{\dagger}_{i}\hat{c}_{i}.
\end{align}
where $\hat{c}_i^{\dagger} (\hat c_i)$ is the creation (annihilation) operator and $L$ denotes the system size. The first term in $\hat H_1$ is the kinetic energy term. The lattice is subjected to two different kinds of onsite potentials. The second term in $\hat H_1$ corresponds to a quasiperiodic onsite potential, and $\hat H_2$ has a contribution from the onsite potential energy due to the linear gradient field with strength $h$.} Here $\omega$ is taken as $\omega = F_n/F_{n+1}$, where $F_n$ is the $n{\text{th}}$ Fibonacci number. For the $n \to \infty$ limit, $\omega = \lim_{n \to \infty} F_n/F_{n+1} \to (\sqrt{5}-1)/2$, which is the golden ratio. In the presence of the quasiperiodic potential, appropriate scaling emerges for the system sizes belonging to the Fibonacci series  \cite{sahoo2024stark}. $\phi$ is a phase which is chosen randomly from a uniform distribution in [0,1]. We consider an open boundary condition (OBC).

\begin{figure}[t]
    \centering
\includegraphics[width=0.4\textwidth]{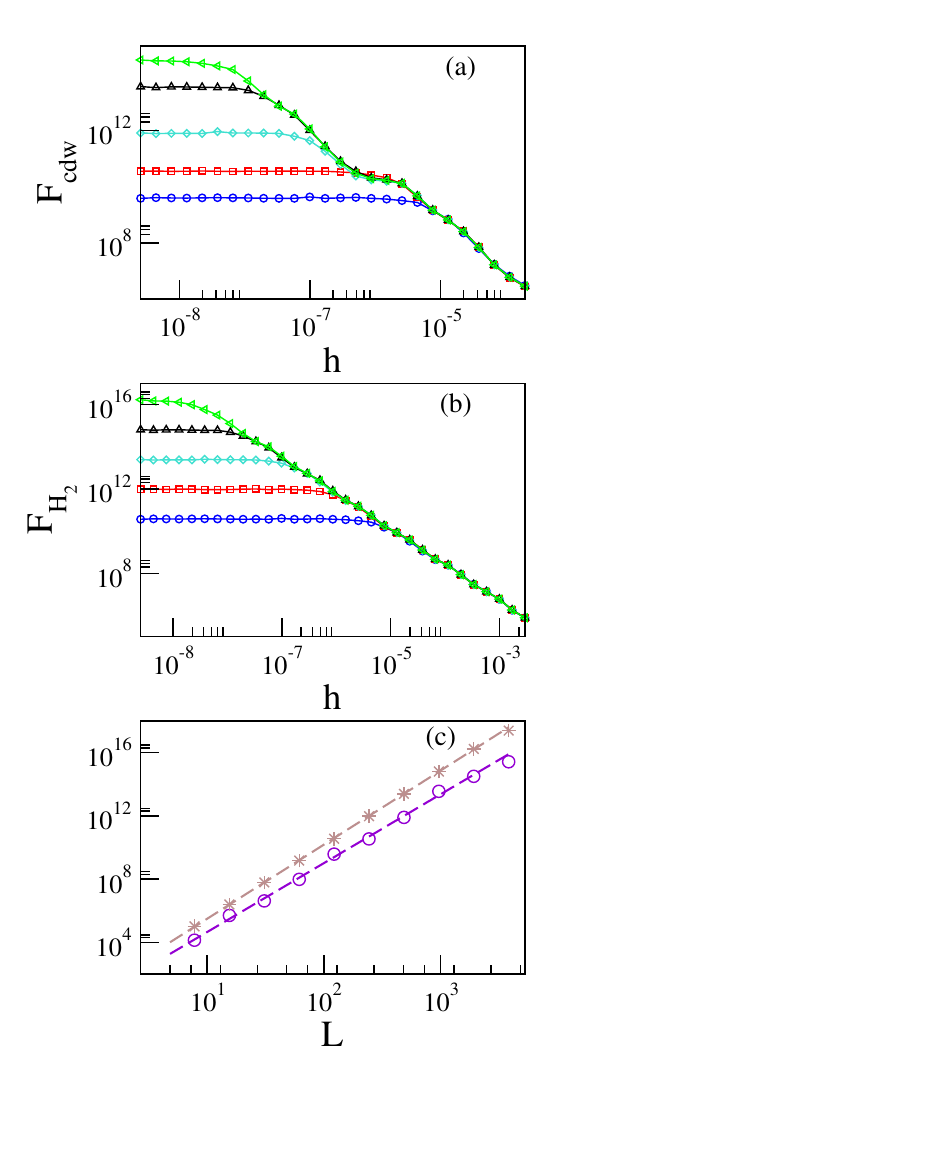}
\caption{{\emph{Single party OFI.}} { (a)} presents OFI, $F_{cdw}$, for observable   $\hat{O}_{cdw}$ with respect to $h$ for system size L = 144 (blue circle), 233 (red square), 377 (turquoise diamond), 610 (black triangular up), 987 (green triangular left).  { (b)}  presents OFI, $F_{H_2}$ for observable   $\hat{O}_{H_2}$ with respect to $h$ for system size L = 144 (blue circle), 233 (red square), 377 (turquoise diamond), 610 (black triangular up), 987 (green triangular left). {(c)} illustrates the scaling of OFI corresponding to the operator $\hat{O}_{cdw}$, $F_{cdw}$ (circles) and the operator $\hat{O}_{H_2}$, $F_{H_2}$ (stars) {\color{black}{for L = 21, 34, 55, 89, 144, 233, 377, 610, 987, 1597  }}. The straight lines are the best fits. Whereas $F_{cdw}$ scales as $F_{cdw} \sim L^{6.2} $, $F_{H_2}$ saturates the scaling of the QFI, i.e., $F_{H_2} \sim L^{6.7} $.  A configuration averaging over $\phi$ is
performed with 8000 random samples.}
 \label{fig:fig2}
\end{figure}

\section{Single-particle ground state as probe}


{\color{black}{{\emph{Quantum Fisher information}. We first consider the single-particle scenario. The Hamiltonian ($\hat H$) reduces to the pure Stark case for $V = 0$ (i.e., in the absence of a quasiperiodic potential). In this limit, the system becomes localized as the field strength approaches zero, $h \to 0$. At this transition point, the fidelity susceptibility scales with system size as $\chi_Q \sim L^{2/\nu}$ with $\nu = 0.33$ for the single-particle ground state, and the QFI exhibits the scaling $F_Q \sim L^{6}$ \cite{PhysRevLett.131.010801}. When the system is additionally tuned to the critical point of the AAH model, i.e., $V = 2$, the Hamiltonian experiences the combined effect of two distinct onsite potentials. In this case, it is known from our previous work  \cite{sahoo2024stark} that the scaling exponent associated with the localization–delocalization transition turns out to be $\nu = 0.29$, and the QFI in the delocalized regime scales as $F_Q \sim L^{\beta}$ with $\beta = 6.7$. 

{\color{black} When the parameter to be estimated is very small, raw Fisher information can appear comparatively more optimistic from an experimental point of view. In such situations, the relative quantum signal-to-noise ratio (RQSNR) provides a complementary metric. To connect QFI with operational sensitivity, we compute the RQSNR defined as $\mathcal{Q}_Q=h\sqrt{MF_Q}  \cite{Paris2009_QuantumEstimation, Mitchison2020_InSituThermometry,Cavazzoni2025_FreqJumps}$, where $M$ is the number of repetitions of the sensing protocol. In practice, the system sizes we analyze are already within reach of current optical-lattice experiments, which have achieved chains as large as of the size of several hundreds \cite{Scherg2021_nonergodicity_tiltedFH, Kohlert2023_fragmentation}. Typical repetition numbers in such experiments are few thousands. In order to provide a concrete example, we consider a particular case with $M=10^3$ and $L=377$, for which we find that for the pure Stark case $\mathcal{Q}=2$, while the introduction of the additional control field in the form of the Stark field in the AAS model enhances the value to $Q=14.7$. {\color{black}{A value of $o(10)$ falls well within the range typically considered sufficient for high-precision metrological applications, especially when estimating weak fields \cite{PhysRevA.111.062602, PhysRevLett.126.020401}}}. The achieved operational sensitivities in the AAS model for estimating weak field is higher compared to the pure Stark model, confirming the metrological advantage caused by the additional control parameter in the form of the quasiperiodic potential considered in this work.} \\

{\emph{Observable Fisher information}. Despite the fact that the QFI continues to provide the ultimate bound on estimation precision, independent of measurement strategy, experimentally accessing the QFI poses significant challenges. In order to overcome this practical constraint, we need to identify an observable that is both experimentally relevant and provides a significant quantum advantage in parameter estimation, which forms the central theme of this work. 

We first focus on the operator $\hat{O}_{cdw} = \sum_i (-1)^i \hat{c}_i^{\dagger}\hat{c}_i$, which measures the occupation imbalance between even and odd sites. This operator reveals the charge-density-wave (CDW) order in a quantum state and can be directly measured in optical lattice setups with ultracold atoms \cite{doi:10.1126/science.aaa7432,PhysRevLett.90.055501,Modugno_2010}.} The computation of OFI is followed the Eq.~\eqref{eq:OFI}, which entails the average of $\langle O_{cdw}\rangle$ and $\langle \hat{O}_{cdw}^2\rangle$, where, $\hat{O}
_{cdw}^2 = \sum_{i,j} (-1)^{i+j} \hat{c}_i^{\dagger}\hat{c}_i\hat{c}_j^{\dagger} \hat{c}_j$.} 
In Fig.~\ref{fig:fig2}(a), we plot OFI, $F_{cdw}$, versus $h$. In the finite-size system for the weak value of $h$, $F_{cdw}$ {\color{black}{has a similar trend as QFI (the QFI plot is shown in Fig. 9 of Ref. \cite{sahoo2024stark}).}} Initially, it remains flat in the extended region, but after a certain value of $h$, $F_{cdw}$ gradually diminishes and it becomes system-size invariant. From this figure,  $F_{cdw}$ increases with increasing system size in the extended region. The scaling of OFI with $L$ is shown in Fig.~\ref{fig:fig2}(c). {\color{black}{The violet circles represent the value of $F_{cdw}$ in the extended region for various system sizes up to 1597, and the dotted violet line is the fitting function $F_{cdw} \sim L^{\beta}$ with $\beta = 6.2$. The scaling thus beats the HL. Setting the control parameter at $V=2$, the scaling exponent of OFI provides high precision for estimation of the weak field.}} Hence, the experimentally realizable CDW observable offers a genuine quantum advantage in the measurement precision. 

Next, we consider another observable $\hat{O}_{H_2} = \sum_i i \hat{c}^\dagger_i\hat{c}_i $, which is essentially the last part of our Hamiltonian. In order to evaluate OFI for the operator $\hat{O}_{H_2}$, we need to evaluate the expectation values of the  operator $\langle \hat{O}_{H_2}\rangle$ and $\langle\hat{O}_{H_2}\rangle^2$, which is basically $\langle\hat{O}_{H_2}\rangle^2 = \sum_{i,j} ij \hat{c}_i^{\dagger}\hat{c}_i\hat{c}_j^{\dagger} \hat{c}_j$. In Fig.~\ref{fig:fig2}(b), we depict OFI, $F_{H_2}$, against $h$ for various system sizes $L$. For a given system size $L$, the OFI, corresponding to $\hat{O}_{H_2}$ has a similar nature to $F_{cdw}$, i.e., initially for a small value of $h$ the $F_{H_2}$ is flat and after a certain value of $h$, it gradually decreases. There, it becomes system size independent. However, the absolute value of OFI for $\hat{O}_{H_2}$ is greater than $F_{cdw}$. In the flat region, the $F_{H_2}$ increases steadily with the system size, $L$. The scaling of $F_{H_2}$ in the extended region is shown in  Fig.~\ref{fig:fig2}(c). {\color{black}{The brown star illustrates the value of $F_{H2}$ at $h=10^{-9}$ with different system sizes $L$ up to 1597,  and the dotted brown line is the fitting function of $F_{H_2} \sim L^{\beta}$ with $\beta = 6.7$.}} This turns out to be a remarkable result as the scaling exponent for OFI  $\hat{O}_{H_2}$ saturates to that of the QFI. \\

{\color{black}{{\emph{Resource analysis by taking account of state preparation time.}} In criticality-based quantum sensing using adiabatic state preparation, it is essential to take into account the preparation time $t$ required to reach the target eigenstate, leading to a modified figure of merit defined by the normalized QFI, $F_Q/t$. Near a quantum critical point, the adiabatic evolution time scales with the system size as $t \sim L^{z}$. The exponent $z$ is related to the closing of the energy gap near criticality via $\Delta E \sim L^{-z}$, where $\Delta E$ denotes the energy difference between the ground and first excited states~\cite{PhysRevX.8.021022}. At the AAH critical point ($V = 2$), the energy gap exhibits the scaling $\Delta E \sim L^{-2.37}$~\cite{sahoo2024stark}, implying $z \approx 2.37$. Incorporating this scaling into the resource cost of adiabatic preparation modifies the QFI scaling from $F_Q \sim L^{\beta}$ to $F_Q/t \sim L^{\beta - z}$. For our case, we find $F_Q/t \sim L^{4.3}$. A similar analysis for the OFI yields $F_{\text{cdw}}/t \sim L^{3.8}$ and $F_{H_2}/t \sim L^{4.3}$. These results demonstrate that, even after accounting for the additional cost of state preparation time, both QFI and OFI retain super-Heisenberg scaling, thereby maintaining a strong quantum advantage over classical sensing limits.}}

\begin{figure}[t]
    \centering
\includegraphics[width=0.5\textwidth]{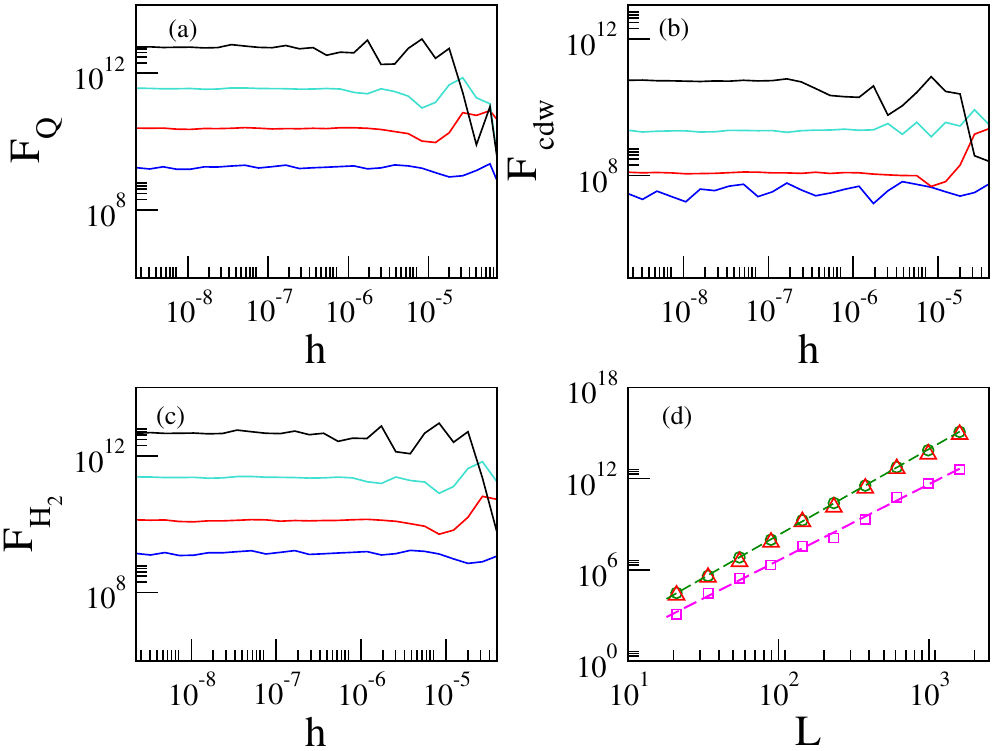}
\caption{{\emph{Single-party mideigenstate.}} {(a)} QFI for the mideigenstate against $h$ for system size $L = 144$ (blue), 233 (red), 377 (turquoise), and 610 (black).  { (b)} presents OFI, $F_{cdw}$, for observable   $\hat{O}_{cdw}$ with respect to $h$ for system size $L = 144$ (blue), 233 (red ), 377 (turquoise), and 610 (black).  {(c)} OFI, $F_{H_2}$ for observable   $\hat{O}_{H_2}$ with respect to $h$ for system size $L = 144$ (blue), 233 (red ), 377 (turquoise), and 610 (black). {(d)} illustrates the scaling of QFI (circular) and OFI, $F_{cdw}$ (square) and $F_{H_2}$ (triangular up) {\color{black}{for $L = 21$, 34, 55, 89, 144, 233, 377, 610, 987, and 1597}}. The straight lines are the best fits. Whereas $F_{cdw}$ scales as $F_{cdw} \sim L^{5} $ shown by the dotted magenta line, $F_{H_2}$ saturates the scaling of the QFI, i.e., $F_{H_2} \sim L^{5.6} $, shown by the dotted green line.  A configuration averaging over $\phi$ is
performed with 8000 random samples.    }
 \label{fig:mid_eigenstate}
\end{figure}

\section{Midspectrum}

{\color{black}{In disordered or tilted (Stark) lattices, the transition between localized and extended behavior does not arise from a single ground-state critical point, but rather from the structure of eigenstates across the spectrum. In the presence of a mobility edge, the localization length diverges as near, the criticality and the rate of divergence is determined by the scaling exponent $\nu$, which is actually energy dependent. This is fundamentally different from a conventional second-order quantum phase transition (QPT), where only the ground state and low-lying excitations capture the QPT directly. In localized systems, each energy window can probe a different effective critical theory, leading to explicitly energy-resolved scaling behavior. Hence, it is interesting to check the effectiveness of the mid-spectrum states as the probe states for high-precision sensing in the localization-based quantum critical sensors. Moreover, the single-particle Stark model can be realized in optical lattices by applying a controlled linear potential, e.g, magnetic field gradients. Transition to an excited state may then be coherently addressed using RF fields, or far-detuned two-photon Raman coupling with negligible spontaneous-emission–induced decoherence.}}

We investigate the QFI and OFI for midspectrum eigenstates, as depicted in Fig.\ref{fig:mid_eigenstate}. Similar to the ground state, the QFI of the midspectrum eigenstate, $F_Q$, exhibits a plateau up to a certain value of the parameter $h$. Beyond this value, $F_Q$ begins to decrease nonuniformly. Within the plateau region, the QFI increases with the system size, as demonstrated for $L = 144$ (blue), $233$ (red), $377$ (Turquoise), and $610$ (green). This plateau region corresponds to the extended phase, where $F_Q$ exhibits a scaling relation $F_Q \sim L^{5.6}$, as shown in Fig. \ref{fig:mid_eigenstate}(a). Although the scaling exponent assumes a lower value in comparison to the ground-state case, it still surpasses the Heisenberg limit for parameter estimation of $h$. In addition, we also analyze the OFI for the CDW operator in the mid-spectrum regime, as shown in Fig. ~\ref{fig:mid_eigenstate}(b). Similar to the QFI, the OFI for the CDW operator remains flat within the extended phase for smaller values of $h$. However, beyond this region, it begins to decrease nonuniformly due to the interplay between the two competing potentials, the AA potential and the Stark potential. Within the flat region, the OFI for the CDW operator scales as $F_{cdw} \sim L^5$, which also exceeds the Heisenberg limit in measurement precision. Furthermore, Fig. \ref{fig:mid_eigenstate}(c) illustrates the OFI for the operator $H_2$, which is found to saturate the QFI for the midspectrum eigenstate. The scaling relationships for these quantities are summarized in Fig. \ref{fig:mid_eigenstate}(d). Here, the green circles represent the QFI, while the red triangles correspond to the OFI of $H_2$, which saturates the QFI. The green dotted line represents the fitting function with the scaling relation $F_Q \sim F_{H_2} \sim L^{5.6}$. Additionally, the magenta squares correspond to the OFI for the CDW operator, with a dotted magenta line indicating the fitting function $F_{cdw}\sim L^5$. These results collectively highlight the superior scaling behavior of both QFI and OFI in the extended phase, surpassing the Heisenberg limit.

\section{Thermal probe}

In the many-body scenario, accessing all individual eigenstates of the system becomes computationally and experimentally challenging due to the vast size of the Hilbert space and the complexity of interactions. In practice, the thermal state is often utilized as a probe. The thermal state is defined as 
\begin{equation}
\rho(h,T) = e^{-H/{k_BT}}/\text{Tr}[e^{-H/{k_BT}}],
 \label{eq:thermal_state}
\end{equation}
where $T$ is the temperature of the system and $k_B$ is the Boltzmann constant. In this context, temperature $T$ introduces an additional parameter that influences the behavior of the system. To quantify the sensitivity of the thermal state we have used the definition of QFI as \cite{PhysRevLett.127.080504}
\begin{equation}
 F_Q (T,h) = \sum_{r,s = 1}^{L} \frac{2\text{Re}(\langle \lambda_r|\partial_h \rho|\lambda_s\rangle \langle \lambda_s|\partial_h \rho|\lambda_r\rangle)}{\lambda_r + \lambda_s},
 \label{eq:mixed_state_QFI}
\end{equation}
where $\rho$ is the density matrix representing the thermal state and can be written in its spectral decomposition as $\rho =  \sum_{r=1}^{L} \lambda_r |\lambda_r\rangle \langle \lambda_r|$. Here $\lambda_r$ are the eigenvalues, and $|\lambda_r\rangle$ are the corresponding eigenvectors of $\rho$. Re(.) represents the real part of the quantity within the parentheses and the sum excludes the term for which $\lambda_r + \lambda_s = 0$.\\

\begin{figure}[t]
    \centering
\includegraphics[width=0.4\textwidth]{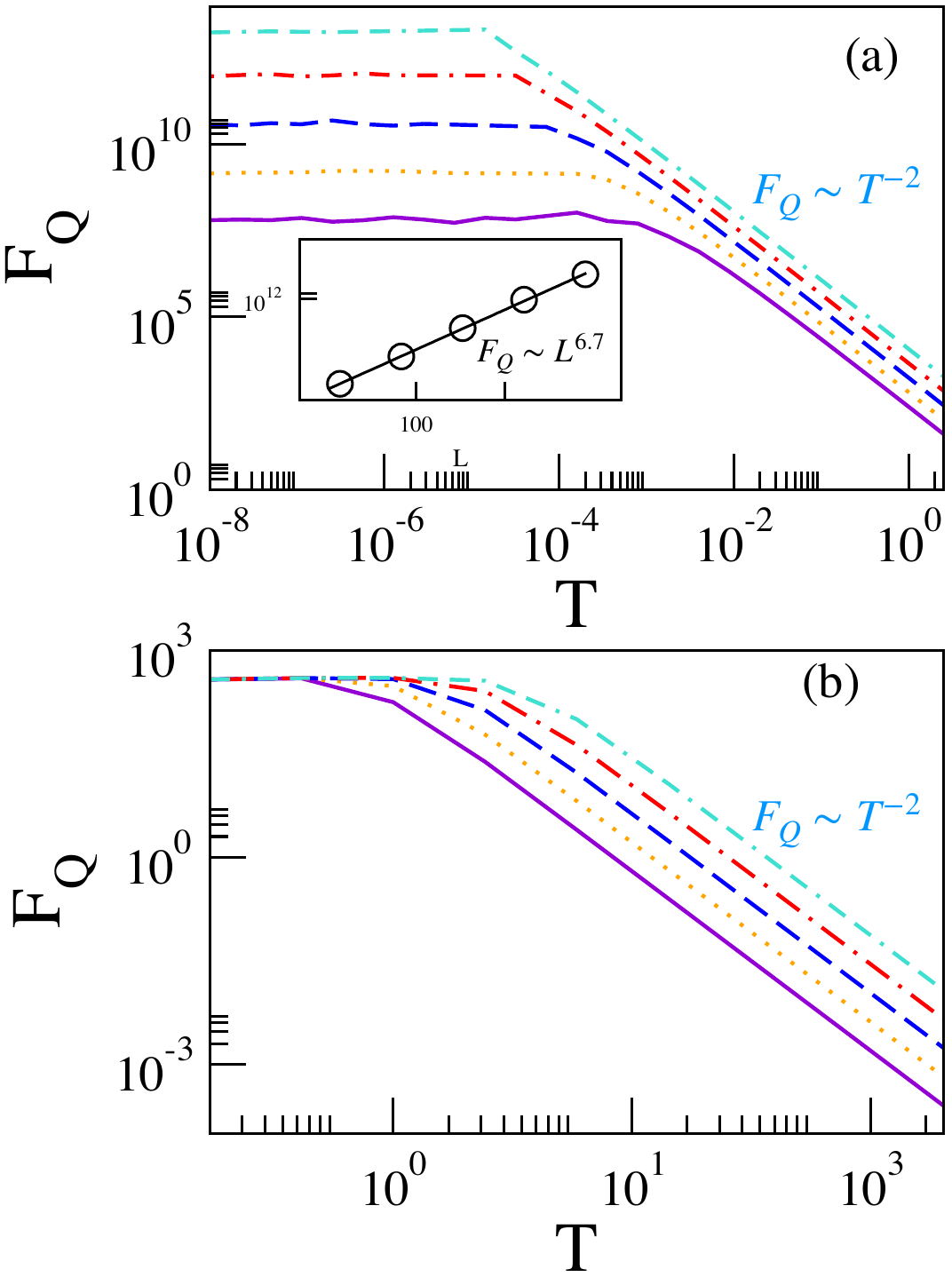}
\caption{{\emph{Single party thermal state probe for QFI.}} {(a)} {\color{black} {presents an analysis of the QFI for the thermal state, $\rho(T,h)$, where $T$ stands for the temperature. Setting $h$ at a low value ($h=10^{-8}$), the $F_Q$ exhibits distinct values for different system sizes at low temperatures. At sufficiently low temperature, the QFI remains flat and recovers the scaling results obtained from the ground state, i.e., $F_Q \sim L^{6.7}$ (see the inset). As the temperature increases, there is a transition in the behavior of $F_Q$, and it is marked with a monotonic decay with the temperature. There, the scaling relation turns out to be $F_Q\sim T^{-2}$.  {(b)} depicts $F_{Q}$ with temperature $T$ when $h$ is set at a comparatively large value, $h=0.05$, where the considered finite-size systems are in the localized phase at the low temperatures. Here at sufficiently low temperatures, all $F_Q$ turns out to be system-size invariant in a similar fashion reported for the ground state. Beyond a transition temperature, $F_Q$ decreases with $T$, $F_Q \sim T^{-2}$. In both cases, the considered system sizes are $L = 55$ (solid), $89$ (dot), $144$ (dash), $233$ (dash-dot), $377$ (dash-dash-dot). A configurational averaing with 500 random realizations of $\phi$ is performed.}}}
 \label{fig:thermal_probe}
\end{figure}

\begin{figure}[t]
    \centering
\includegraphics[width=0.5\textwidth]{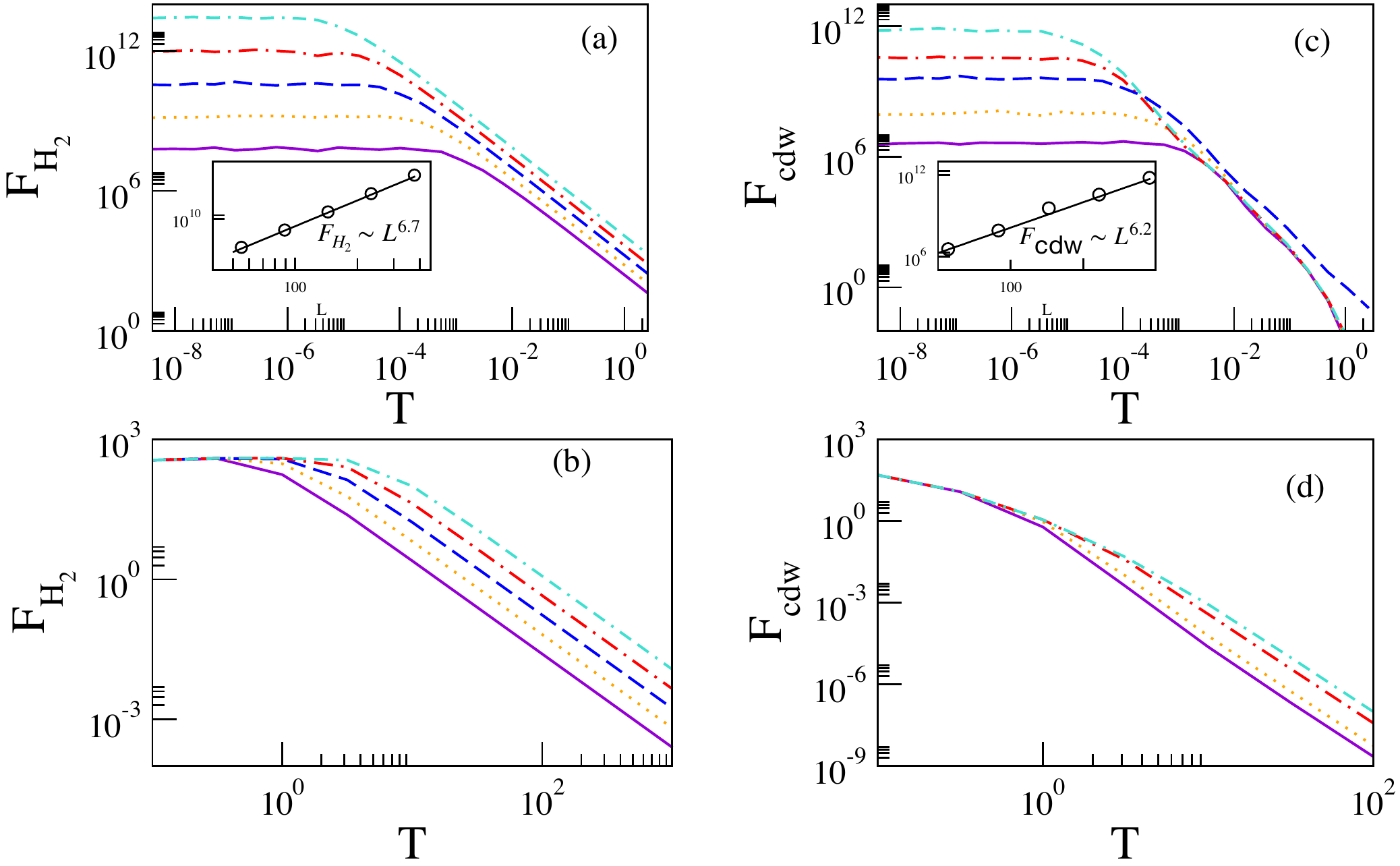}
\caption{{\emph{Single party thermal state probe for OFI.}} {\color{black} { {(a)} presents behavior of the OFI, $F_{H_2}$, corresponding to the operator, $\hat{O}_{H_2}$ in the thermal state $\rho(T,h)$, for $h = 10^{-8}$. OFI has a trend similar to QFI (See. Fig.~3), and the scaling relation here is $F_{H_2} \sim L^{6.7}$ (see inset) at sufficiently low-temperatures, similar to the behavior observed in the ground state. As the temperature increases, $F_{H_2}$ transitions to a decay regime, following the scaling relation $F_{H_2} \sim T^{-2}$.  {(b)} Temperature dependence of $F_{H_2}$ in the localized phase ($h = 0.05$). At low temperatures, $F_{H_2}$ collapses across system sizes, resembling ground-state behavior. Beyond a certain temperature, $F_{H_2}$ decays with $T$, while maintaining the scaling, $F_{H_2} \sim T^{-2}$. {(c)} For $h = 10^{-8}$), $F_{\text{cdw}}$ also follows the ground-state scaling relation, $F_{\text{cdw}} \sim L^{6.2}$ (see inset), at sufficiently low temperatures as well. Above a certain temperature, however, $F_{\text{cdw}}$ decreases with increasing $T$. {(d)} In the localized phase ($h = 0.05$), the temperature dependence of $F_{\text{cdw}}$ shows a collapse across system sizes at low temperatures, consistent with ground-state behavior. At higher temperatures, $F_{\text{cdw}}$ decreases with increasing $T$.  In all the cases, considered system sizes are $L = 55$ (solid), $89$ (dot), $144$ (dash), $233$ (dash-dot) and $377$ (dot-dash-dot). We have performed an average over 500 random values of $\phi$.}} }
 \label{fig:thermal_probe_OFI}
\end{figure}

We compute the QFI with respect to the parameter $h$, while temperature $T$ is treated as the varying parameter. To explore the system's behavior under different conditions, we consider two distinct scenarios, one where the system is in the extended phase at sufficiently low temperatures, for which we set the gradient field as characterized by $h=10^{-8}$, and another where it is in the localized phase, the chosen value for which is $h=0.05$. At low temperatures, specifically when the thermal energy scales as $k_BT$, and becomes smaller than the energy gap in the eigenenergy spectrum, 
$\Delta E$, i.e, $k_BT<\Delta E$, the system predominantly behaves like its ground state. This occurs because the ground state possesses the maximum Gibbs probability in the thermal distribution. For $h=10^{-8}$, the sufficiently low-temperature regime exhibits a distinct separation of QFI curves for different system sizes $L$ as depicted in Fig.~\ref{fig:thermal_probe}(a). {\color{black}{In this regime, the QFI inherits the same scaling behavior as the ground state in the extended phase, i.e., $F_Q \sim L^{6.7}$.}} In contrast, when the system resides in the localized phase ($h=0.05$), the low-temperature QFI plots collapse onto a single curve, irrespective of system size, shown in  Fig.~\ref{fig:thermal_probe}(b). This collapse is characteristic of the localized systems for the ground state. \\

For higher temperatures, when $k_BT\ge\Delta E$, the system displays intriguing behavior in both the extended and localized phases. As shown in Fig.~\ref{fig:thermal_probe}, the QFI monotonically decays as a function of temperature. The QFI decays with temperature as $F_Q \sim T^{-2}$. Notably, despite the temperature-induced decay, the QFI in this region continues to scale as $F_Q \sim L^{2}$, thereby saturating the Heisenberg limit. This scaling saturation implies that the system's quantum feature remains robust even at elevated temperatures. To encapsulate this behavior, the QFI can be universally described by the scaling relation

\begin{equation}
 F_Q (T,h) \sim f(h)T^{-2}L^2,
 \label{eq:thermal_scaling}
\end{equation}
{\color{black} {where $f(h)$ is an arbitrary function of $h$.}} This universal scaling relation offers a comprehensive framework for understanding the interplay between $T$, $h$, and $L$ in determining the QFI. \\

 We also analyze the OFI corresponding to the operator $\hat{O}_{H_2}$ for thermal states and observe that its behavior closely resembles that of the QFI with respect to temperature at sufficiently low temperatures. In Fig.~\ref{fig:thermal_probe_OFI}(a), we present the variation of $F_{H_2}$ as a function of temperature $T$ for $h = 10^{-8}$, which corresponds to the extended phase of the system. At very low temperatures, $F_{H_2}$ plots exhibit distinct separations for different system sizes $L$. This separation is consistent with the behavior previously observed for the ground state, and the scaling in this regime is $F_{H_2} \sim L^{6.7}$, indicating a consistent behavior when compared to the ground state case.  As the temperature increases, the OFI undergoes a power-law decay following the relation $F_{H_2} \sim T^{-2}$. In this relatively high-temperature regime, however, the system exhibits a different scaling behavior-- the OFI grows with system size as $F_{H_2} \sim L^2$, indicating the HL. This highlights the robustness of Heisenberg scaling for thermal states at high temperatures in the extended phase.

In Fig.~\ref{fig:thermal_probe_OFI}(b), we examine the behavior of $F_{H_2}$ in the localized phase, setting $h = 0.05$. At low temperatures, $F_{H_2}$ for different system sizes collapses onto a single universal curve. This collapse reflects the results in the localized phase corresponding to the ground state case. As the temperature increases, $F_{H_2}$ again decays monotonically, following the same power-law relation $F_{H_2} \sim T^{-2}$ as in the extended phase. In the high-temperature regime, despite being in the localized phase, the OFI still shows scaling with system size as $F_{H_2} \sim L^2$, once again reaching the HL. 

The behavior of the OFI-corresponding CDW operator is presented in Fig.~\ref{fig:thermal_probe_OFI}(c). In the delocalized phase, $F_{\text{cdw}}$ increases with system size $L$, exhibiting a scaling exponent of $6.2$, which coincides with that of the ground state. This indicates that at low temperatures, $F_{\text{cdw}}$ retains the same scaling behavior as in the ground state. As the temperature increases, however, $F_{\text{cdw}}$ begins to decrease with $T$, and in this high-temperature regime, the scaling with $L$ is no longer observed. In contrast, in the localized phase [Fig.~\ref{fig:thermal_probe_OFI}(d)], $F_{\text{cdw}}$ shows a collapse across system sizes at low temperatures, again consistent with ground-state behavior. Upon further increasing the temperature, $F_{\text{cdw}}$ decreases monotonically with $T$, exhibiting HL scaling with system size.

Thus, for relatively large temperatures, the behavior of the OFI across both phases can be captured by a unified scaling ansatz: $F_{H_2}(T, h) \sim f(h) T^{-2} L^2$, where $f(h)$ is an arbitrary function of $h$. This scaling behavior emphasizes that the value of $h$ strongly affects the low-temperature response of the system.  In the relatively high-temperature regime, despite the fact that the system is dominated by thermal fluctuations, we find the Heisenberg-limited scaling with respect to the system size. However, it is worth emphasizing, as that is expected, the absolute values of QFI and OFIs in the high-temperature regime are lower than in the low-temperature regime, implying higher effectiveness of the probes at lower temperatures. For this system, we obtain a quantum advantage in estimating the parameter at low temperatures. As the temperature increases, the advantage in measurement precision gradually decreases. Nevertheless, it is important to note that this trend is not universal. Several alternative models demonstrate that, under specific conditions, high-temperature regimes can actually provide greater precision than their low-temperature counterparts. In such cases, thermal excitations may enhance sensitivity to the parameter of interest \cite{Ostermann2024, mihailescu2025_singular, Garc2024}.

\begin{figure}[t]
    \centering
\includegraphics[width=0.4\textwidth]{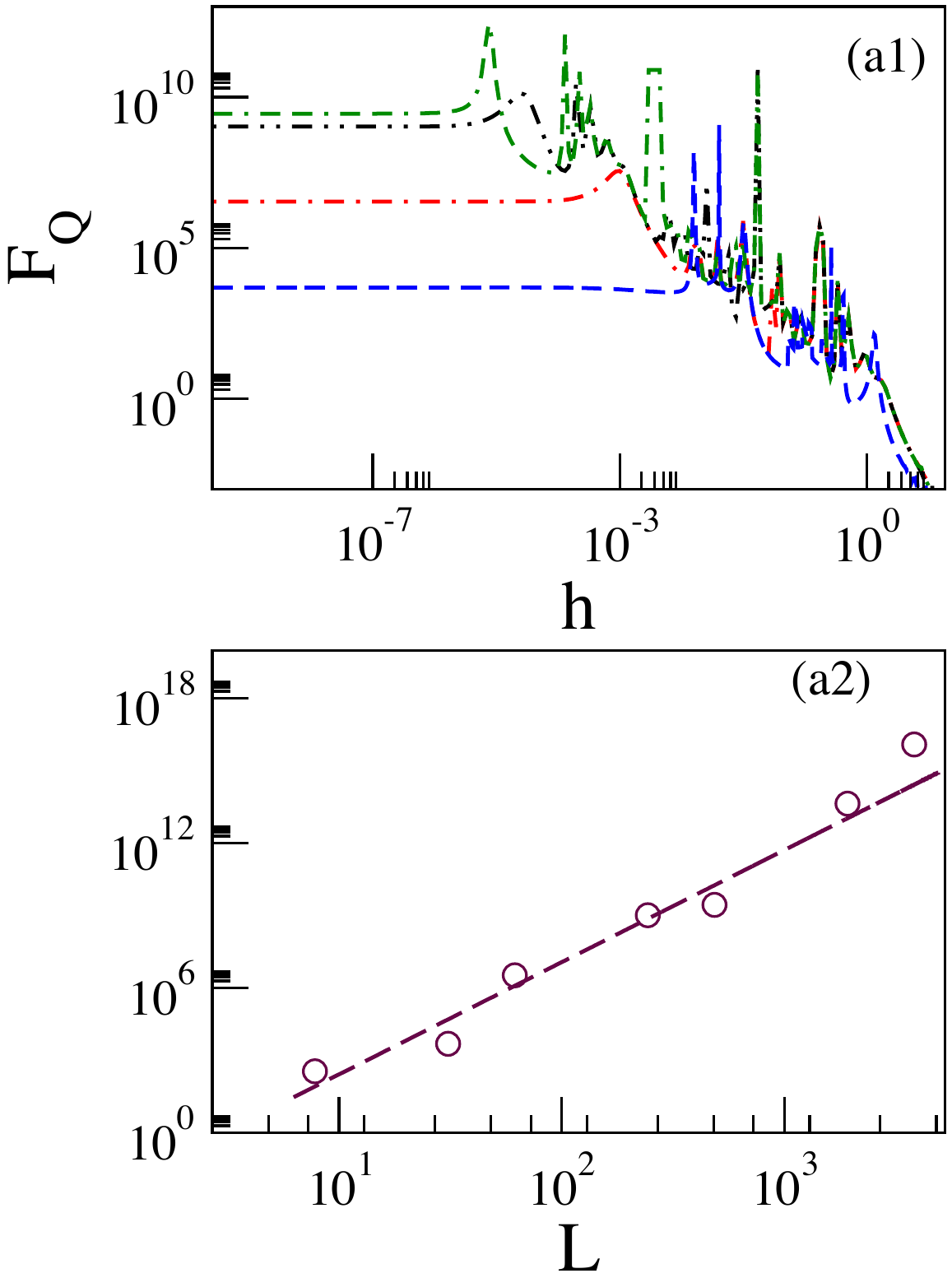}
\caption{{\emph{Half-filled case QFI.}}  {(a1)} QFI for half filled case with respect to $h$. We have taken system size $L (n_f)$, where $L$ is the system size and $n_f$ is the number of filling. This figure shows that in the flat region, the QFI is increasing with system sizes, $L$ = 55 (28) (blue dash-dash), 89 (44) (red dot-dash), 233 (116) (black dot-dot-dash), 377 (188) (green dash-dash-dot). After a certain value of $h$, the QFI decays with certain initial fluctuations, and again it shows steady behavior.
  {(a2)} In this plot, we have shown the scaling of QFI with system size $L (n_f)$. The circle point is numerical value of QFI for $h = 10^{-9}$ for $L$ = 21 (10), 55 (28), 89 (44), 233 (116), 377 (188), 987 (493), {\color{black}{1597(798)}} and the dotted line is the best fitting with fitting function $F_Q(h=10^{-9}) \sim L^{6.6}$.}
 \label{fig:fig3}
\end{figure}

\begin{figure}[t]
    \centering
\includegraphics[width=0.4\textwidth]{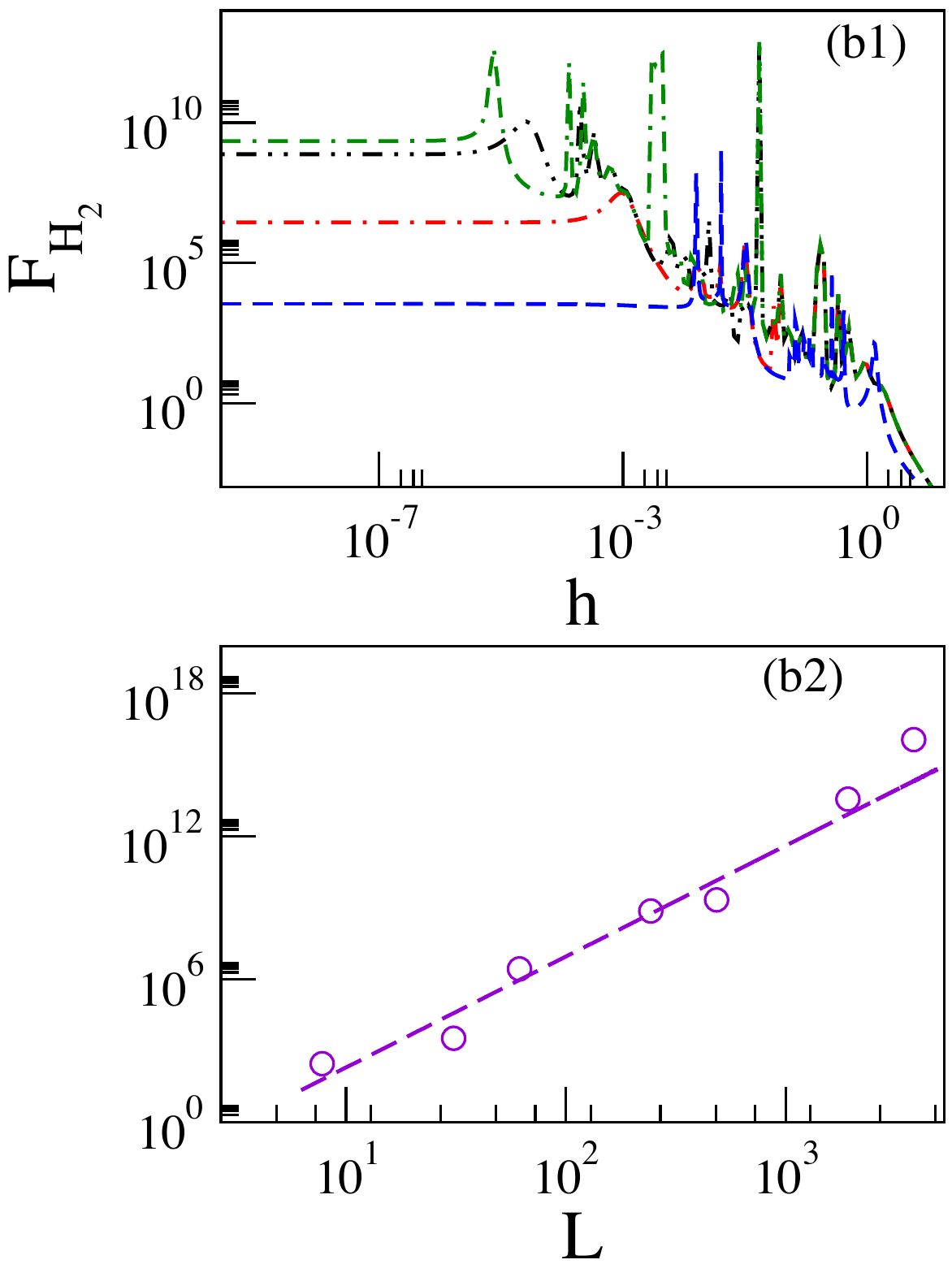}
\caption{{\emph{Half-filled case OFI.}} {(b1)} The figure illustrates the behavior of the OFI for system size $L (n_f)$, where $n_f$ is the number of fermions, with respect to $h$. In the flat region, the OFI increases with system size. However, after reaching a certain value of $h$, the OFI begins to decay with fluctuations and eventually stabilizes into a steady behavior. We have taken $L$ = 55 (blue dash-dash), 89 (red dot-dash), 233 (black dot-dot-dash), 377 ( green dash-dash-dot).  {(b2)} In this plot, we show the scaling of OFI with system size. The violet circular points represent the numerical values of OFI for  $h = 10^{-9}$ with $L$ = 21 (10), 55 (28), 89 (44), 233 (116), 377 (188), 987 (493), {\color{black}{1597(798)}}, while the violet dotted line represents the best fit. We obtain $F_{H_2}(h=10^{-9}) \sim L^{6.6}$.}
 \label{fig:fig4}
\end{figure}

\section{Half-filled system as probe }
We now focus on the half-filled case. We consider odd lattice sizes with OBC and the corresponding number of particles is either one of these two options: $n_f = (L \pm 1)/2$. We perform a slater determination method for this half-filled case. In order to construct the ground-state configuration, we diagonalize the single-particle Hamiltonian and take the first $n_f$ eigenvectors in the form of Slater determination. The ground state for the half-filled case is given by,
\begin{equation}
\psi_0(j_1,......,j_{n_f}) = \frac{1}{\sqrt{n_f!}}
\begin{vmatrix}
\psi_1(j_1) & ... & \psi_1(j_{n_f})\\
... & ... & ...\\
... & ... & ...\\
\psi_{n_f}(j_1) & ... & \psi_{n_f}(j_{n_f})\\
\end{vmatrix} ,
\end{equation}
where $j$ runs from 1 to $n_f$. The fidelity for the half-filled ground state $\psi_0$ is 
\begin{equation}
f = |\langle \psi_0 (h)|\psi_0 (h+dh) \rangle|.    
\end{equation}
The general form of fidelity between $\psi_0(h)$ and $\psi_0(h+dh)$ is
\begin{equation}
f =|\langle \psi_0(h)|\psi_0(h+dh)\rangle| = 
\begin{vmatrix}
P_{11} & ... & P_{1n_f}\\
... & ... & ...\\
... & ... & ...\\
P_{n_f1} & ... & P_{n_fn_
f}\\
\end{vmatrix} ,
\end{equation}
where $P_{lm} = |\langle\psi_l(h)|\psi_m(h+dh)\rangle|$. $\psi_l, \psi_m$ are $l\text{'th}$ and $m\text{'th}$ eigenvectors. Now after taylor series expansion, the fidelity can be written as 
\begin{equation}
f = |\langle\psi_0(h)|\psi_0(h+dh)\rangle| = 1 - \frac{(\delta h)^2}{2} \chi   , 
\end{equation}
where $\chi = -\langle \psi_0|\frac{d^2\psi_0}{dh^2} \rangle$. The fidelity susceptibility $\chi = 2 (1 - f)/dh^2$ and the corresponding QFI is $F_Q = 4 \chi$. 

We compute the ground state $F_Q$ and $F_{H_2}$ for system sizes up to 987 ($n_f = 493$), where $n_f$ is number of fermions. For the half-filled probe, we abstain from performing an averaging over $\phi$. We set $\phi = 0$. 
In the following, we describe the results that were obtained.
Fig.~\ref{fig:fig3}(a1) illustrates the QFI for varying $h$. In a finite-size system, within the weak field region, where $h$ is small, the QFI remains constant, exhibiting steady behavior. However, beyond a certain value of $h$, the QFI begins to fluctuate within a specific range due to the trade-off between AAH and the Stark strength. Eventually, its value decreases and stabilizes again. For the weak value of $h$ in a finite-size system, the $F_Q$ increases with system size in the flat region of $F_Q$. So, in this region, there is a scaling of $F_Q$ with fitting function $F_Q \sim L^{\beta}$, where we find the scaling exponent $\beta = 6.6$. This is shown in Fig.~\ref{fig:fig3}(a2) with the dotted maroon line, and the maroon circles denote the numerical value of $F_Q$ at $h=10^{-9}$. The half-filled Stark probe also achieves the super-Heisenberg limit with $\beta = 6.6$.

  The operator $\hat{O}_{cdw}$ is not ideal because the ground state lacks CDW ordering in the localized phase. This means that, unlike in the single-particle scenario, the OFI for $\hat{O}_{cdw}$ does not exhibit proper scaling with $L$. However, the observable $\hat{O}_{H_2} = \sum_i i \hat{c}_i^{\dagger}\hat{c}_i$ does provide an appropriate scaling. The corresponding OFI for the finite size system is shown in Fig.~\ref{fig:fig4}(b1). This figure illustrates the OFI as a function of $h$. Initially, for small values of $h$, the OFI remains relatively constant. However, beyond a certain threshold of $h$, the OFI starts to fluctuate. The fluctuations dissipate gradually and eventually stabilize again, similar to the behavior of the QFI. Additionally, the value of the OFI nearly converges to that of the QFI. In the flat region where OFI is not changing with $h$, the scaling of OFI for $h=10^{-9}$ is shown in Fig.~\ref{fig:fig4}(b2). The violet circle is the value of OFI at $h=10^{-9}$, and the dotted violet line is the fitting function of $F_{H_2} \sim L^{\beta}$, where $\beta$ is 6.6. The scaling of OFI for observable $\hat{O}_{H_2}$ is also saturated with the scaling of QFI. From Fig.~\ref{fig:fig4}(a1) and Fig.~\ref{fig:fig4}(b1), it is clear that the scaling sustains over an extended region, which is an added advantage over many quantum critical sensors for which quantum advantage is found over a very narrow region near the quantum criticality. For the half-filled system as a probe, the scaling of QFI beats the Heisenberg limit and gives the super Heisenberg scaling of 6.6. Also, for the half-filled probe, we get an observable that gives a super-Heisenberg scaling of $\beta = 6.6$ saturating the scaling of QFI.

\section{Conclusions} In the realm of quantum many-body systems, achieving super-Heisenberg limit precision remains a formidable challenge. This paper focuses on engineering efficient quantum many-body sensors. In an excellent recent new work it has been proposed that efficient weak-field sensors can be designed by utilizing Stark localization as a resource \cite{PhysRevLett.131.010801}. Specifically, it has been demonstrated that under Stark localization for ground state probe, the quantum Fisher information (QFI) scales as $L^6$ and for midspectrum state the QFI scales as $L^{4.1}$, i.e., a beyond-Heisenberg-limit scaling is achieved. 

This work proposes to use different classes of localization-inducing potentials that can enhance the precision of parameter estimation even further. Near the criticality of the AAH potential, better precision in estimating weak fields of Stark strength can be achieved with the QFI scaling as $L^{6.7}$ for ground state, and $L^{5.6}$ for the midspectrum state, respectively. In comparison, for the pure Stark system, the QFI is reported to scale as $L^{6}$ and $L^{4.1}$ for ground state and midspectrum state, respectively. In addition, this work examines the charge-density-wave (CDW) order, $\hat{O}_{cdw}$, which is an experimentally accessible observable. This observable also exhibits super-Heisenberg scaling with the OFI scaling as $F_{cdw} \sim L^{6.2}$ for ground state as a probe. Moreover, another observable, $\hat O_{H_2}$, has been identified for which the associated OFI not only saturates to the QFI but also matches the scaling behavior of the QFI. For the midspectrum state, the OFI for CDW scales with $L$ as $F_{cdw} \sim L^{5}$, and the scaling exponent of OFI for $\hat O_{H_2}$ saturates the scaling of the QFI. Now, for the thermal state as a probe, we have shown that the QFI saturates to the HL for all ranges of the temperature considered in this work when the system is prepared in the delocalized phase ($h=10^{-8}$). When the system is at localized phase $h = 0.05$, the QFI collapses for all system sizes at the low temperature limit, i.e., the system remains scale invariant and there is no scaling with system sizes. Still, at higher temperatures the QFI saturates to the HL. However, as expected, the absolute values of the QFI and OFI in the high-temperature regime are reduced compared to those at low temperatures, indicating that the probes are more effective when thermal fluctuations are minimal.} Further studies have been performed for the system at half-filling, for which the scaling of the QFI with system size for the ground state is found to be $F_Q \sim L^{6.6}$. Similarly, for the half-filled case, the OFI corresponding to the operator $\hat O_{H_2}$ exhibits the same scaling behavior as the QFI, that is $F_{H_2} \sim L^{6.6}$.\\ 

In summary,  our work proposes an experimentally realizable protocol for achieving enhanced sensing of the weak Stark field by invoking another type of localization-inducing potential in the form of the AAH potential and by exploiting the AAH criticality. One may, in principle, envision the underlying strength of the concept which can, in principle, be generalized for criticality-based quantum sensors -- In the context of precision estimation of an unknown parameter within the QMB platform, it might be advantageous to introduce additional control parameters, that, in turn, introduces new kinds of quantum criticality, for obtaining enhanced quantum advantage by utilizing the competing effects of multiple critical points.


\bibliography{References}

\end{document}